\documentclass[preprint2]{aastex6}

\usepackage{natbib}
\usepackage{amsmath}
\usepackage{braket}
\begin{document}
\def\vec#1{\mbox{\boldmath $#1$}}

\title{High-frequency spicule oscillations generated via mode conversion}

\author{Munehito Shoda\altaffilmark{1} and Takaaki Yokoyama\altaffilmark{2}}
\affil{Department of Earth and Planetary Science, The University of Tokyo \\
Hongo, Bunkyo-ku, Tokyo, 113-0033, Japan}

\altaffiltext{1}{shoda@eps.s.u-tokyo.ac.jp}

\begin{abstract}

	Spicule oscillations involve high-frequency components with a typical period approximately corresponding to $40-50$ s.
	The typical time scale of the photospheric oscillation is a few minutes, and thus, the origin of this high-frequency component is not trivial.
	In this study, a one-dimensional numerical simulation is performed to demonstrate that the observed spicule oscillations originate from longitudinal-to-transverse mode conversion 
	that occurs around the equipartition layer in the chromosphere.
	Calculations are conducted in a self-consistent manner with the exception of additional heating to maintain coronal temperature.
	The analyses indicate the following features:
	(1) mode conversion efficiently excites high-frequency transverse waves; 
	(2) the typical period of the high-frequency waves corresponds to the sound-crossing time of the mode conversion region; and
	(3) simulated root-mean-square velocity of the high-frequency component is consistent with the observed value.
	These results indicate that the observation of spicule oscillation provides direct evidence of mode conversion in the chromosphere.
		
\end{abstract}

\keywords{magnetohydrodynamics (MHD) --
methods: numerical -- Sun: chromosphere -- Sun: oscillations}

\section{Introduction} \label{sec:introduction}

	In order to sustain the chromospheric and coronal temperature at an observed level, it is necessary to continuously supply energy \citep{Withb77,Ander89}.
	Various types of waves are emitted from the surface magneto-convection \citep{Alfve47,Oster61,Balle98,Stein98,Chitt12,Kato016,Oba0017}, 
	and thus, the wave heating scenario is frequently examined as a candidate of solar atmospheric heating.
	
	Longitudinal waves are generated by the vertical motions of the photosphere \citep{Leigh62,Khome01,Kato011,Kato016,Oba0017}.
	In fact, several observations indicate the existence of the longitudinal waves in the chromosphere \citep{Mein076,Tian014,Kanoh16}.
	From a theoretical viewpoint, the radiating shock waves that evolve from longitudinal waves can explain the observed feature of chromospheric spectral line profiles \citep{Carls92,Carls97}, 
	and this also indicates the chromospheric longitudinal waves.
	The longitudinal waves more or less contribute to the chromospheric heating. 
	However, they do not supply sufficient energy into the corona because most of the energy flux of longitudinal waves is consumed inside the chromosphere \citep{Mein081,Kanoh16}.
	Therefore, transverse waves are likely to play a role in the corona and solar wind.
	
	Transverse waves as well as longitudinal waves are excited on the photosphere \citep{Kulsr55,Sprui82,Edwin83,Morto13} 
	either by the swaying motion of the flux tube \citep{Stein98} or the vortex motion inside the flux tube \citep{Balle98,Balle11,Iijim17}.
	It is shown that transverse waves can transport sufficient energy into the corona both theoretically \citep{Kudoh99,Cranm05} and observationally \citep{DePon07a,McInt11,Thurg14}.
	Specifically, incompressible (Alfv\'en) waves are directly observed in the solar wind \citep{Colem68,Belch71a}. 
	Furthermore, several coronal heating \citep{Hollw86,Suzuk05,Cranm07,Verdi10,Shoda18a} and solar wind acceleration \citep{Belch71b,Jacqu77,Heine80} models
	successfully explain the observation based on Alfv\'en wave modeling.
	Therefore, it is important to investigate the generation, propagation, and dissipation of transverse waves to clarify the energy budget inside the corona and solar wind.
		
	The chromospheric jets are preferable targets of transverse wave observation.
	With respect to the spicules  \citep{Becke68,DePon07b,DePon11},  
	{\it Hinode} \citep{Kosug07} observation revealed that a sufficiently high amount of energy is transported into the corona \citep{DePon07a,Suema08,Perei12,Ebadi12}.
	Ground-based observation obtains a similar quantity of transverse waves \citep{Jess012}.
	The studies indicate that the transverse motion involves typical velocity amplitude of $15-25 {\rm \ km\ s^{-1}}$ and typical period of $150-350 {\rm \ s}$.
	The observed amplitudes are lower for fibrils and mottles although this could be due to observational constraints \citep{Morto14}.
	{\it SDO}/AIA \citep{Lemen12} observation of transverse waves in the transition region and corona \citep{McInt11,Thurg14} is consistent with the spicule observation, 
	while the ground-based observation by CoMP \citep{Tomcz08} results in a significantly lower amplitude \citep{Tomcz07}, and this could be due to the superposition along the line of sight.
	Both the chromospheric and coronal observation reveal that the typical period of transverse motion corresponds to a few minutes, and this is comparable to the time scale of granular motion \citep{Matsu10b}.

	Interestingly, the detailed analysis of spicule oscillation shows that they also involve sub-minute scale waves \citep{He00009,Okamo11} 
	with a typical amplitude of $7-8 {\rm \ km \ s^{-1}}$ and a typical period of $40-50 {\rm \ s}$.
	The photospheric horizontal flow does not exhibit strong power in sub-minute scale \citep{Matsu10b}, and thus, the origin of this high-frequency transverse waves is unclear.
	This could come from the fine-scale vortex motions inside the flux tubes \citep{Balle98,Balle11,Chitt12}, although it is unlikely that these motions will generate the observed swaying motion of spicules.
	Therefore, it is possible that high-frequency waves are generated in the interface region between the photosphere and transition region, i.e., the chromosphere.
		
	The plasma and magnetic field are highly inhomogeneous, and the plasma beta is approximately unity in the chromosphere.
	Hence, this is a preferable region for waves to couple with each other \citep{Rosen02,Bogda03}.
	Longitudinal-to-transverse mode conversion occurs near the equipartition region where the sound and Alfv\'en speeds coincide and is typical of these types of coupling processes.
	Both analytical \citep{Schun06,Cally08} and numerical \citep{Fedun11,Khome12,Santa15} studies show that the mode conversion occurs in the chromosphere.
	Several observations also indicate the signature of mode conversion \citep{Jess012,Morto15}.
	
	In this study, we propose a model for the generation of high-frequency spicule oscillations based on the mode conversion.
	The mode conversion is more efficient for higher-frequency waves \citep{Schun06,Cally08}. Thus, it is highly likely that high-frequency transverse waves are likely to appear in spicules \citep{He00009,Okamo11}.
		
	The remainder of this study is organized as follows.
	In Section \ref{sec:method}, we discuss the basic equations and numerical setup.
	The calculation results and analysis are discussed in Section \ref{sec:result}, and we summarize the study in Section \ref{sec:summary}.

\section{Method}  \label{sec:method}

\subsection{Basic equations and set-up}

The basic equations are as follows:
\begin{align}
	&\frac{\partial}{\partial t} \left( \rho A \right) + \frac{\partial }{\partial z} \left( \rho v_z A \right) =0, \label{eq:mass} \\
  	&\frac{\partial}{\partial t} \left( \rho v_z A \right) + \frac{\partial }{\partial z} \left[ \left( \rho {v_z}^2 + p + \frac{{\vec{B}_{\perp}}^2}{8\pi} \right) A \right] \nonumber \\
  	&= \left( p + \frac{\rho {\vec{v}_{\perp}}^2}{2} \right) \frac{dA}{dz}  - \rho g A, \label{eq:eomz} \\
  	&\frac{\partial}{\partial t} \left( \rho \vec{v}_{\perp} A^{3/2} \right) + \frac{\partial }{\partial z} \left[ \left( \rho v_z \vec{v}_{\perp} - \frac{B_z \vec{B}_{\perp}}{4 \pi} \right) A^{3/2} \right]  = 0, \\
  	&\frac{\partial}{\partial t} \left( \sqrt{A} \vec{B}_{\perp} \right) + \frac{\partial }{\partial z} \left[ \left( \vec{B}_{\perp} v_z - B_z \vec{v}_{\perp} \right) \sqrt{A} \right] = 0, \\
    	&\frac{\partial}{\partial t} \left[ \left( e + \frac{1}{2} \rho \vec{v}^2 + \frac{\vec{B}^2}{8 \pi} \right) A \right] \nonumber \\
  	&+ \frac{\partial }{\partial z} \left[\left( e + p + \frac{1}{2} \rho \vec{v}^2  + \frac{{\vec{B}_{\perp}}^2}{4 \pi} \right ) v_z A - B_z \frac{\vec{B}_{\perp} \cdot \vec{v}_{\perp}}{4 \pi} A \right] \nonumber \\
 	&= - L_{\rm rad} A - \frac{\partial }{\partial z} \left( q_{\rm cond} A \right)- \rho g v_z A, \label{eq:energy} \\
	& e = \frac{1}{\gamma-1} p \label{eq:eos}
\end{align}
A generalized form of a spherical coordinate system is used such that the super radial expansion of a flux tube is considered \citep{Hollw82a,Suzuk05} (see Appendix for derivation).
The $xy$ plane is defined as perpendicular to the flux tube while the $z$ axis is curved along the flux tube.
Specifically, $A$ denotes the cross section of the flux tube that satisfies the divergence-free condition of a magnetic field as follows:
\begin{align}
	B_z A = {\rm const.} \label{eq:solenoidal}
\end{align}
$g = 2.74 \times 10^4 {\rm \ cm \ s^{-2}}$ is the gravitational acceleration, $\gamma = 5/3$ corresponds to the specific heat ratio of the adiabatic gas,
$L_{\rm rad}$ is the radiative cooling, and $q_{\rm cond}$ denotes the thermal conductive flux.

Following \citet{Kopp076} and \citet{Suzuk05}, the chromospheric flux tube expansion is modeled as
\begin{align}
  A(z) =  \frac{A_{\rm max} \exp \left( \frac{z - z_1}{\sigma_1} \right) + A_1}{\exp \left( \frac{z - z_1}{\sigma_1} \right)+1},
\end{align}
where
\begin{align}
	A_1&=1-(A_{\rm max}-1) \exp \left( - \frac{z_1}{\sigma_1} \right). 
\end{align}
In this study, we apply $A_{\rm max}=20$ and $z_1 = \sigma_1 = 1 {\rm \ Mm}$.
As for thermal conduction, Spitzer-H\"arm-type flux is employed \citep{Spitz53}.
\begin{align}
	q_{\rm cond} = - \kappa_0 T^{5/2} \frac{\partial T}{\partial z},
\end{align}
where $\kappa_0=10^{-6}$ in the CGS--Gaussian unit.
Approximated cooling functions are included for both optically thick and thin radiations as follows:
\begin{align}
	L_{\rm rad} = (1-\xi) L_{\rm thick} + \xi L_{\rm thin},
	\label{eq:rad}
\end{align}
where $\xi$ denotes the coefficient that controls the contribution of each cooling function.
The chromosphere is dominated by optically thick cooling while the corona is dominated by optically thin cooling, and thus, $\xi$ is set as follows:
\begin{align}
	\xi = \exp \left[ - \frac{\rho}{\rho_{\rm tr}} \right],
	\label{eq:rad_xi}
\end{align}
where $\rho_{\rm tr} = 10^{-14} {\rm \ g \ cm^{-3}}$ denotes mass density near the transition region.
Following \citet{Gudik05}, as opposed to directly solving the radiative transfer \citep{Carls97,Guerr13}, we approximate the optically thick cooling by Newtonian cooling.
This is formulated as
\begin{align}
	L_{\rm thick} = \frac{1}{\tau_{\rm thick}} \left( e - e_0 \right),
\end{align}
where $\tau_{\rm thick}$ is a time scale of the cooling, and $e_0$ denotes an internal energy distribution with a reference temperature model.
$\tau_{\rm thick}$ denotes a function of density that is given as
\begin{align}
	\tau_{\rm thick} = 1.0 \times \left( \frac{\rho}{\rho_\odot} \right)^{-0.4} \ {\rm sec},
\end{align}
where $\rho_\odot$ is the mass density at the photosphere.
With respect to $L_{\rm thin}$, the following expression is used:
\begin{align}
  L_{\rm thin} =  n_i n_e \Lambda (T),
  \label{eq:rad_thin}
\end{align}
where $\Lambda (T)$ is an approximated radiative loss function \citep{Suthe93,Matsu14}.
Additionally, $n_i$ and $n_e$ are calculated by assuming a certain ionization degree as a function of temperature.

\begin{figure*}[!t]
	\begin{center}	 
 	 \includegraphics[width=180mm]{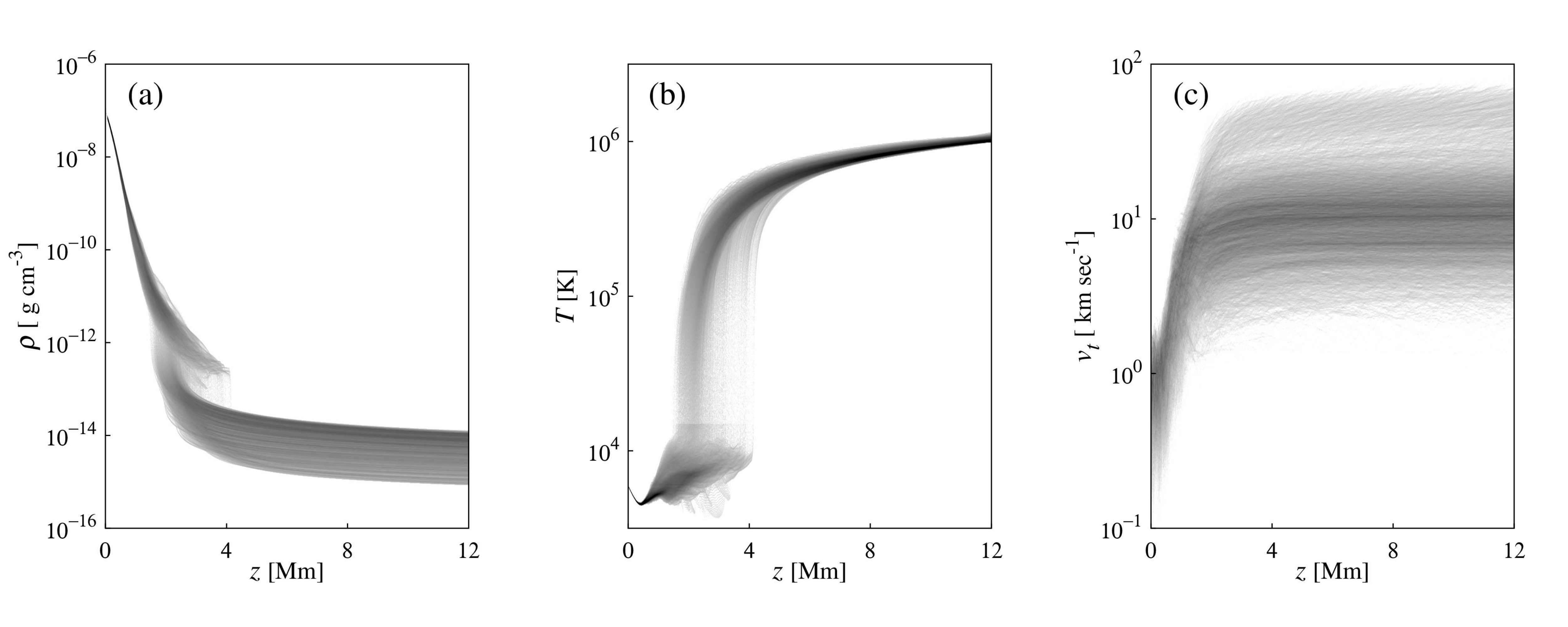}
  	\end{center}
  	\vspace{-1em}
  	\caption{
	Probability distribution function (pdf) of (a) mass density, (b) temperature, and (c) transverse velocity as a function of height.
	Dark colors denote high probability with a logarithmic color table.
  	}
  	\label{fig:pdf}
	\vspace{1em}
\end{figure*}

\subsection{Photospheric boundary condition}

Photospheric flux tube intensity $B_{z,\odot}$ is fixed to $200 {\rm \ G}$.
Radial velocity $v_{z,\odot}$ is given as a monochromatic function in time, while mass density $\rho_\odot$ is determined such that upward waves are excited on the photosphere as follows:
\begin{align}
	v_{z,\odot} = \sqrt{2} \delta v \sin \left( 2 \pi f_0 t \right), \ \ \rho_{\odot} = 10^{-7} \left( 1+\frac{v_{z,\odot}}{c_\odot} \right) {\rm \ g \ cm^{-3}},
\end{align}
where $\delta v$ denotes the photospheric root-mean-square velocity of $v_z$, $f_0$ denotes the input frequency of longitudinal waves, and $c_\odot$ denotes the sound speed at the photosphere.
We set $\delta v$ and $f_0$ to $0.4 {\rm \ km \ s^{-1}}$ and $5 {\rm \ mHz}$, respectively.
Furthermore, $f_0$ is set close to the most dominant frequency of the chromospheric longitudinal waves \citep{Tian014,Kanoh16}.

The transverse velocity fluctuations $\vec{v}_{\perp,\odot}$ are assumed to pose a broadband spectrum while the transverse magnetic field $\vec{B}_{\perp,\odot}$ is given such that the downward Els\"asser variables \citep{Elsas50} disappear at the photosphere. The expression is as follows:
\begin{align}
	\vec{v}_{\perp,\odot} &\propto  \int^{f_{\rm max}}_{f_{\rm min}} f^{-1} e^{2 \pi i f} df, \nonumber \\
	\vec{B}_{\perp,\odot} &= - \sqrt{4 \pi \rho_\odot} \vec{v}_{\perp,\odot},
\end{align}
where $f_{\rm min} = 1{\rm \ mHz}$ and $f_{\rm max} = 10 {\rm \ mHz}$, and the root-mean-square velocity of each component is $0.4 {\rm \ km \ s^{-1}}$.
The vanishing downward Els\"asser variables result from the brevity of numerical calculation.
Several previous studies show the standing waves on the photosphere \citep{Fujim09,Kanoh16} while \citet{Morto11} indicates that the upward propagating mode potentially explains the observation.
The numerical result should be independent of the boundary condition as long as a sufficient energy is injected into the atmosphere.

\subsection{Numerical method}

Basic equations (\ref{eq:mass})-(\ref{eq:eos}) are solved from the photosphere ($z=0{\rm \ Mm}$) to the corona ($z=12{\rm \ Mm}$).
We use 2400 uniform grid points to resolve the computational domain. 
The outgoing (transmitting) boundary condition \citep{DelZa01,Suzuk06a} is applied for the top boundary such that unphysical wave reflection is excluded.
Furthermore, an additional heating is imposed near the top boundary to maintain the coronal temperature \citep{Iijim15}.
The HLLD approximated Riemann solver \citep{Miyos05} is used to solve nonlinear wave propagation.
5th-order accurate WENOZ scheme \citep{Borge08} is used to reduce the numerical dissipation, while third-order SSP Runge--Kutta method \citep{Shu0088} is used for time integration.
The super-time-stepping method \citep{Meyer12,Meyer14} is used to solve the thermal conduction, and this significantly reduces the numerical costs.

\section{Results and Discussion}  \label{sec:result}

\begin{figure}[!t]
	\begin{flushleft}	 
 	 \includegraphics[width=85mm]{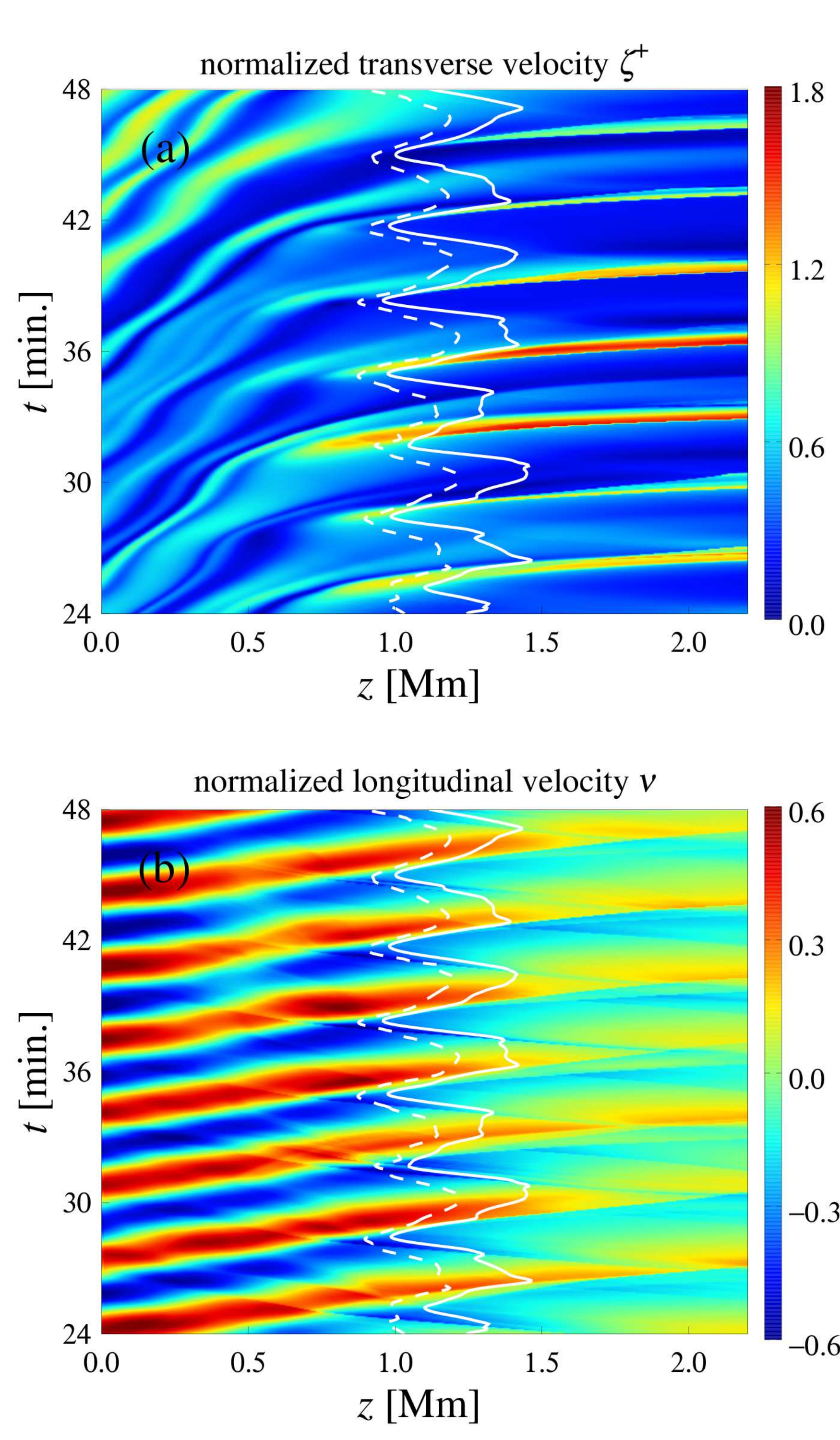} 
  	\end{flushleft}
  	\vspace{-1em}
  	\caption{
		Space--time plots of (a) normalized transverse velocity $\zeta^{+}$ and (b) longitudinal velocity $\nu$ (see eq. (\ref{eq:zeta}) and eq. (\ref{eq:nu}) for definitions) in the chromosphere.
		The unit in the color bar is in ${\rm km \ s^{-1}}$.
		White lines represent the equipartition (conversion) layer.
		The solid line is obtained with adiabatic sound speed $c=\sqrt{\gamma p/ \rho}$ while the dashed line is obtained with isothermal sound speed $c=\sqrt{p / \rho}$.
		  		}
  	\label{fig:space_time}
\end{figure}

\subsection{Wave dynamics}

In Figure \ref{fig:pdf}, we use a probability distribution function (pdf) to show the structure of the chromosphere and the corona in a quasi-steady state.
The mass density, temperature, and transverse velocity are plotted as functions of height.
Dark colors denote high probability with a logarithmic color table.
The obtained density, temperature, and wave amplitude are sufficiently realistic to discuss the chromospheric wave dynamics.

Wave amplitudes should be normalized with respect to the wave action to discuss the energy flux variation.
With respect to the WKB approximation, wave action conservation is expressed as \citep{Breth68}
\begin{align}
	\frac{d}{dt} \left( \frac{E}{\omega^{'}} \right) + \left( \nabla \cdot \vec{c} \right) \left( \frac{E}{\omega^{'}} \right) = 0,
\end{align}
where $E$ denotes the wave energy, $\vec{c}$ denotes the group velocity, and $d/dt=\partial/\partial t + \vec{c} \cdot \nabla$ denotes the material time derivative.
$\omega^{'}$ is the intrinsic frequency that is defined as
\begin{align}
	\omega^{'} = \omega - \vec{k} \cdot \vec{U},
\end{align}
where $\omega$ and $\vec{k}$ denote the wave frequency and wave number, and $\vec{U}$ denotes the mean flow.
In our study, mean flow is negligible $(\vec{U} \approx 0)$, and therefore, $\omega^{'} = \omega$ is a constant.
The action conservation then yields
\begin{align}
	 \frac{\partial}{\partial t} E + \nabla \cdot \left( \vec{F}_{\rm wave} \right) = 0, \ \ \ \ \vec{F}_{\rm wave} = E \vec{c},
\end{align}
where $\vec{F}_{\rm wave}$ denotes the wave energy flux.
Specifically, in a quasi-steady state, $\partial E / \partial t \approx 0$ and
\begin{align}
	F_{\rm wave} A = {\rm const.} \label{eq:action}
\end{align}
The energy flux of upward transverse and longitudinal waves are as follows:
\begin{align}
	&F_{\rm tran} = \frac{1}{4} \rho {\vec{z}_\perp^{+}}^2 a, \label{eq:flux_tran} \\
	&F_{\rm long} = \frac{1}{2} \rho v_z^2 c, \label{eq:flux_long}
\end{align}
where $\vec{z}_{\perp}^{\pm} = \vec{v}_{\perp} \mp \vec{B}_\perp / \sqrt{4\pi\rho}$ denote Els\"asser variables and 
$a = B_z / \sqrt{4\pi \rho}$ and $c = \sqrt{\gamma p /\rho}$ denote the Alfv\'en and sound speed, respectively.
It should be noted that the Els\"asser variables are characteristic variables of Alfv\'en waves in incompressible plasma \citep{Elsas50}.
They are not exact characteristic variables in compressible plasma. However, they are expected to yield a good approximation of transverse wave amplitude \citep{Marsc87,Stone08}.
Eq.s (\ref{eq:action}), (\ref{eq:flux_tran}) and (\ref{eq:flux_long}) yield
\begin{align}
	&\frac{1}{4} \rho {\vec{z}_\perp^{+}}^2 \ a A = {\rm const.}, \label{eq:consflux_tran} \\
	&\frac{1}{2} \rho {v_z}^2 c A= {\rm const.} \label{eq:consflux_long}
\end{align}
The conservation of normalized Els\"asser variable $\zeta^{+}$ is derived from Eq. (\ref{eq:consflux_tran}) as follows:
\begin{align}
	{\zeta^{+}}^2 &= {\rm const.} \ \ \ \ {\rm where} \nonumber \\
	\zeta^{+} &= \left({\frac{\rho}{\rho_\odot}}\right)^{1/4} \sqrt{{\vec{z}_{\perp}^{\pm}}^2} ,
	\label{eq:zeta}
\end{align}
while the normalized longitudinal velocity $\nu$ is derived from Eq. (\ref{eq:consflux_long}) as follows:
\begin{align}
	\nu^2 &= {\rm const.} \ \ \ \ {\rm where} \nonumber \\
	\nu &= \left( \frac{\rho }{\rho_\odot} \right)^{1/4} \left( \frac{p }{p_\odot} \right)^{1/4} \left( \frac{B_z }{B_{z,\odot}} \right)^{-1/2} v_z.
	\label{eq:nu}
\end{align}
Dissipation and reflection that are neglected in the WKB approximation decrease $\zeta^{+}$ and $\nu$ with respect to the height.
$\zeta^{+}$ and $\nu$ increase only when wave energy supply exists.

Figure \ref{fig:space_time} shows the space--time plot of $\zeta^{+}$ and $\nu$.
The solid white line represents the equipartition layer with adiabatic sound speed ($c_{\rm adi}=\sqrt{\gamma p / \rho}$),
while the dotted white denotes the equipartition layer with isothermal sound speed ($c_{\rm iso}=\sqrt{p / \rho}$).
The sound speed ranges between $c_{\rm adi}$ and $c_{\rm iso}$ based on the timescale of the Newtonian cooling and wave period, and thus the equipartition layer lies between the solid and dotted lines.
$\zeta^{+}$ is clearly amplified near the equipartition layer, and this is especially evident when high $\nu$ exists near the white line.
This directly implies that longitudinal-to-transverse mode conversion occurs near the equipartition layer, 
and this is consistent with the results of previous studies \citep{Bogda03,Schun06,Khome12}.
The most important parameter for mode conversion, the angle between wave vector and magnetic field line (the attacking angle $\alpha$), is zero in the absence of transverse waves 
because waves are assumed to propagate along the background field line in our system.
The mode conversion observed in the simulation is triggered by wave--wave interaction, 
and thus the efficiency of each mode conversion event is never predictable because $\alpha$ changes relative to time based on the amplitude and phase of the transverse wave.
Hence, the pdf of the transverse velocity (Fig. \ref{fig:pdf} (c)) is vertically broadened when compared with the density and temperature.
This differs from the results of the previous studies that consider the interaction between waves and background magnetic fields.

\begin{figure}[!t]
	\vspace{1em}
	\begin{flushleft}	 
 	 \includegraphics[width=78mm]{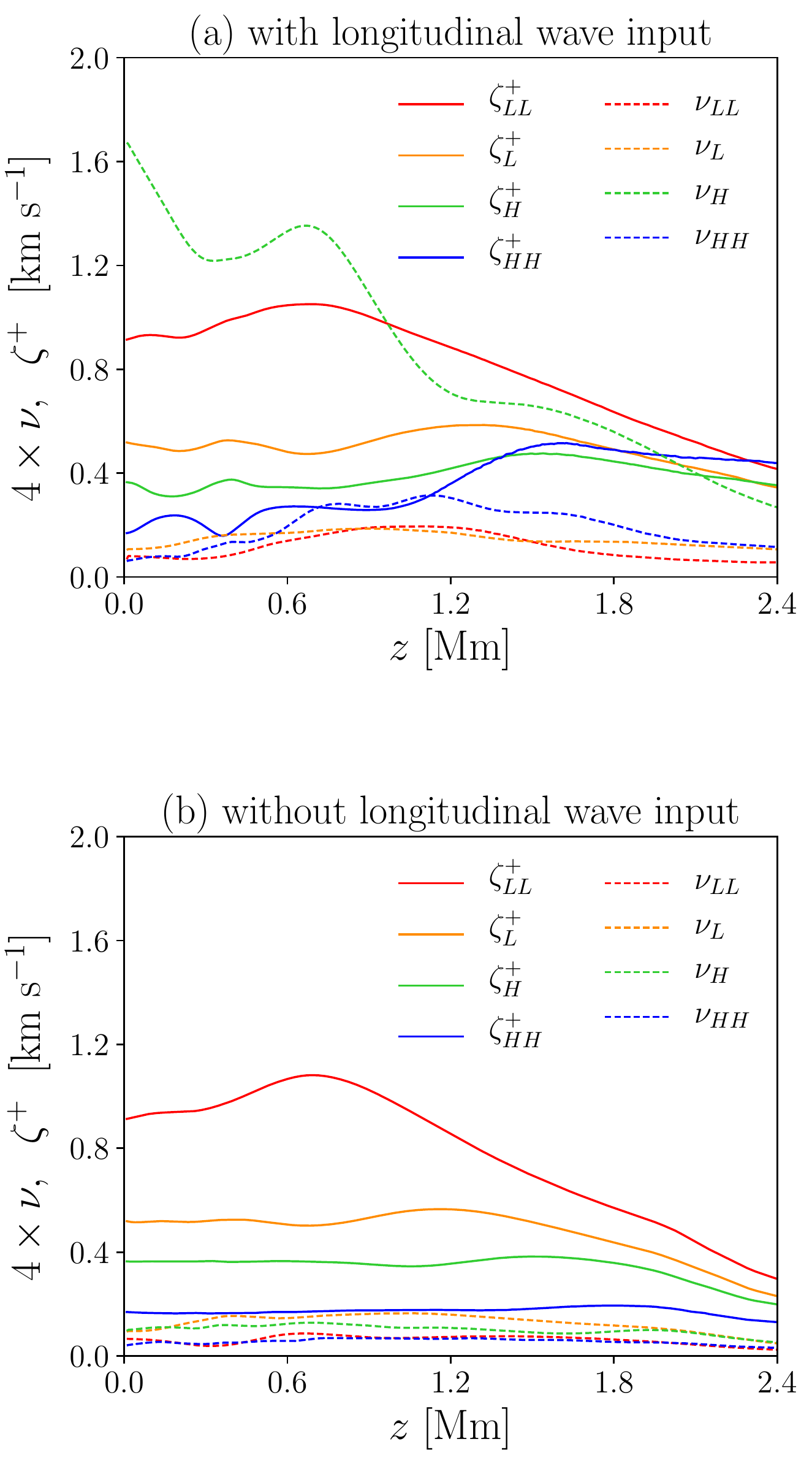} 
  	\end{flushleft}
  	\vspace{-1em}
  	\caption{
			Frequency-decomposed, normalized Els\"asser variable $\zeta^{+}$ (solid lines), and longitudinal velocity $\nu$ (dashed lines) as functions of height.
			$4 \times \nu$ is shown instead of $\nu$ for clearer description. 
			Upper (a) and lower (b) panels show the cases with and without longitudinal wave excitation on the photosphere, respectively.
			Red, orange, green, and blue lines indicate 
			very-low-frequency ($\zeta^+_{LL}, \ \nu_{LL}$), low-frequency ($\zeta^+_{L}, \ \nu_{L}$), 
			high-frequency ($\zeta^+_{H}, \ \nu_{H}$), and very-high-frequency ($\zeta^+_{HH}, \ \nu_{HH}$) components, respectively
			(see Eq.s (\ref{eq:fourier}) and (\ref{eq:decompose}) for definitions).
		  		}
  	\label{fig:zeta_freq}
\end{figure}

\subsection{Frequency decomposition} \label{sec:freq_decomp}

In order to perform a detailed analysis, we applied frequency decomposition into each normalized variable.
With respect to arbitrary variable $\eta(z,t)$, the decomposition is conducted in the following procedure.
First, the Fourier transformation is applied for each $z$ as
\begin{align}
	\tilde{\eta} (z,\omega) = \frac{1}{2\pi T} \int^T_0 \eta (z,t) \exp(-i \omega t) dt,
	\label{eq:fourier}
\end{align}
where $T=480$ minutes, corresponding to the total simulation time.
From $\tilde{\eta} (z,\omega)$, the decomposed values are calculated as follows:
\begin{align}
	&\eta_{LL} (z)   =  \sqrt{\int_{|\omega|<2 \pi f_1} |\tilde{\eta} (z,\omega)|^2 \ d\omega}, \nonumber \\
	&\eta_{L} (z)     =  \sqrt{\int_{2\pi f_1<|\omega|<2 \pi f_2} |\tilde{\eta} (z,\omega)|^2 \ d\omega}, \nonumber 
\end{align}
\begin{align}
	&\eta_{H} (z)    =  \sqrt{\int_{2 \pi f_2<|\omega|<2 \pi f_3} |\tilde{\eta} (z,\omega)|^2 \ d\omega}, \nonumber \\
	&\eta_{HH} (z)  = \sqrt{\int_{2 \pi f_3<|\omega|} |\tilde{\eta} (z,\omega)|^2 \ d\omega},
	\label{eq:decompose}
\end{align} 
where $f_1 = 2.5 {\rm \ mHz}$, $f_2 = 5 {\rm \ mHz}$, and $f_3 = 10 {\rm \ mHz}$.
We refer to $\eta_{HH}$, $\eta_{H}$, $\eta_{L}$, and $\eta_{LL}$ as 
the very-high-frequency component, high-frequency component, low-frequency component, and very-low-frequency component, respectively.
The normalized Els\"asser variable $\zeta^{+}$ and longitudinal velocity $\nu$ are decomposed in this manner.

\begin{figure}[!t]
	\vspace{1em}
	\begin{flushleft}	 
 	 \includegraphics[width=90mm]{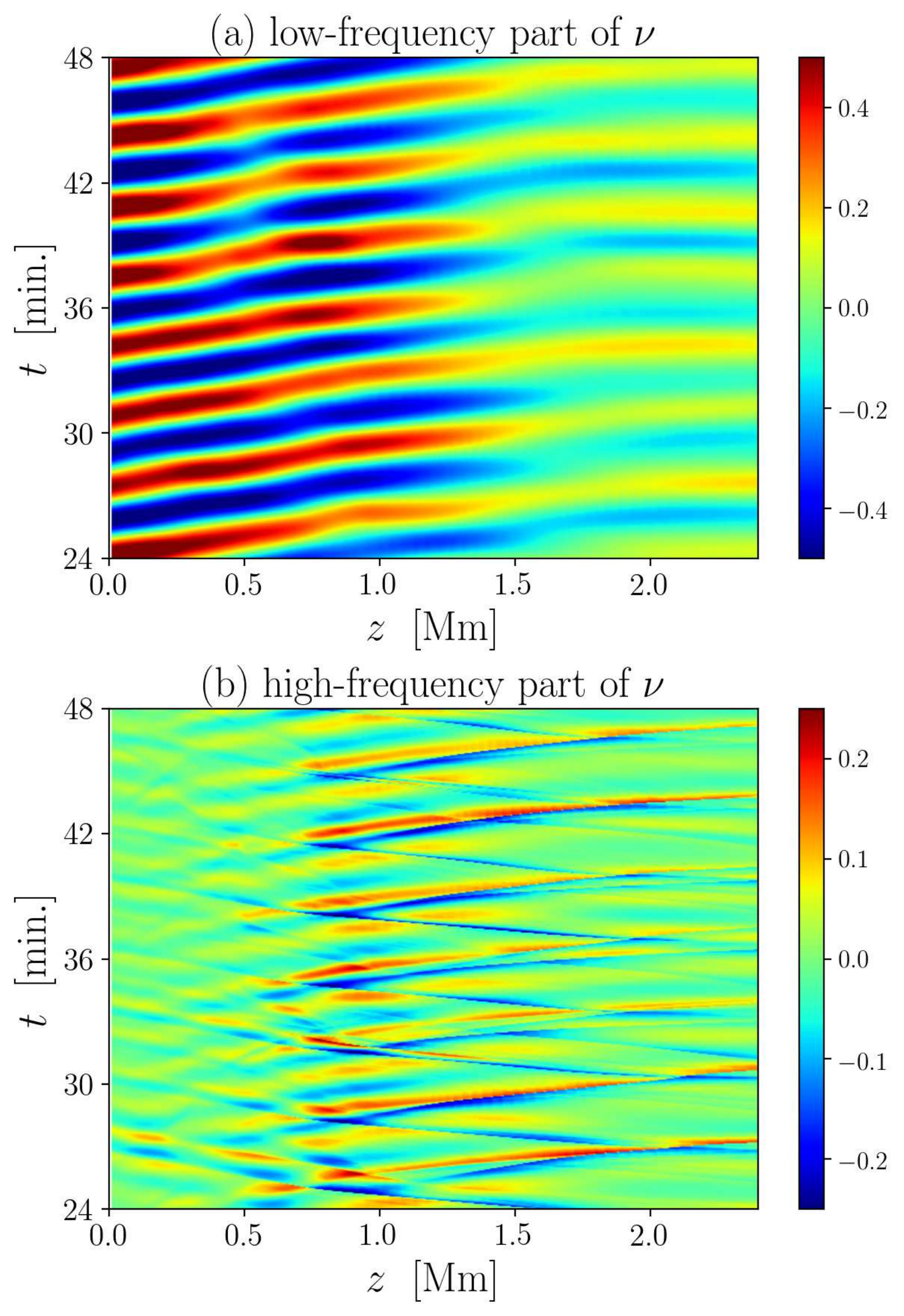} 
  	\end{flushleft}
  	\vspace{-1em}
  	\caption{
			Space-time plots of frequency filtered $\nu$ with cut-off frequency of $10 {\rm \ mHz}$.
			Panel (a) and (b) correspond to low-frequency and high-frequency parts, respectively.
			Units in the colorbar is ${\rm km \ s^{-1}}$.
		  		}
  	\label{fig:nu_space_time}
\end{figure}

\begin{figure*}[!t]
	\begin{center}	 
 	 \includegraphics[width=140mm]{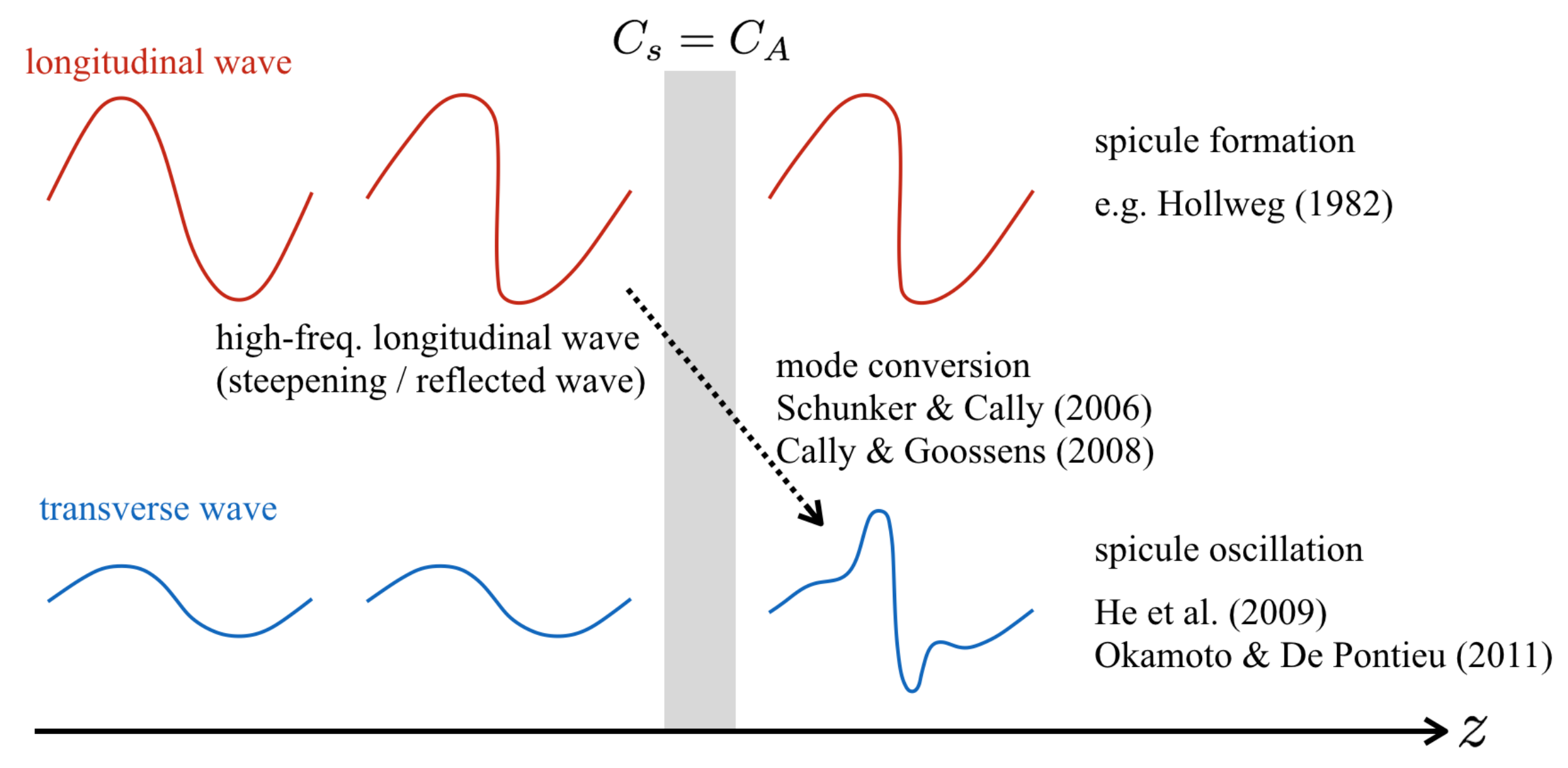}
  	\end{center}
  	\vspace{2em}
  	\caption{
	Schematic picture of the physical process in our simulation.
  	}
  	\label{fig:sum}
	\vspace{0em}
\end{figure*}

\begin{figure*}[!t]
	\begin{center}	 
 	 \includegraphics[width=160mm]{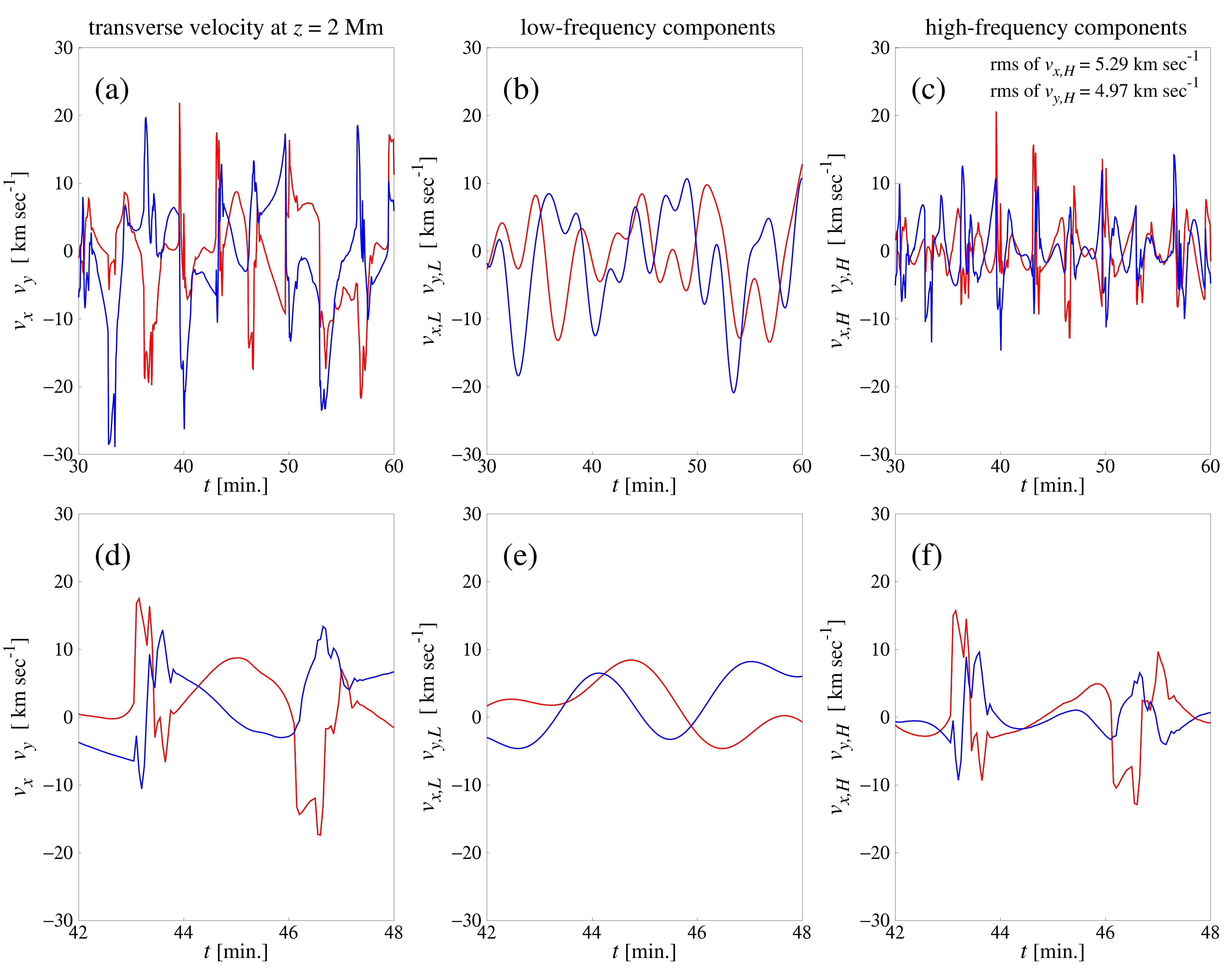}
  	\end{center}
  	\vspace{2em}
  	\caption{
	Transverse velocities at $z=2{\rm \ Mm}$ (a) and their low-frequency (b) and high-frequency (c) components.
	Red and blue lines indicate $v_x$ and $v_y$, respectively.
	Lower panels show the same quantities with a shorter temporal range.
  	}
  	\label{fig:vt}
	\vspace{0em}
\end{figure*}

Figure \ref{fig:zeta_freq} shows the decomposed $\zeta^{+}$ (solid line) and $4 \times \nu$ (dashed line) as functions of height.
It should be noted that we focus on $4 \times \nu$ as opposed to $\nu$ to provide a better description.
Red, orange, green, and blue lines correspond to the
very-high-frequency component, high-frequency component, low-frequency component, and very-low-frequency component, respectively.
In order to demonstrate the role of mode conversion, the results with (upper panel) and without (lower panel) longitudinal wave inputs are simultaneously shown.
The lower panel shows a natural character of Alfv\'en wave propagation in which the low-frequency mode experiences reflection \citep{Velli93,Cranm05,Verdi07} 
while the high-frequency mode conserves its energy flux \citep{Heine80}.
The profiles of the low-frequency components (red and orange lines) in upper and lower panels are similar to each other, and this suggests that they are not influenced by the longitudinal waves
while the high-frequency waves are amplified in the upper panel.
This behavior is consistent with the mode conversion scenario because higher mode conversion rate is observed (transmission rate is smaller) for high-frequency waves \citep{Schun06,Cally08}.
In fact, $\zeta^{+}_{HH}$ increases when $\nu_{HH}$ decreases in $z=1.2-1.8{\rm \ Mm}$, and this demonstrates the energy transport from $\nu$ to $\zeta^{+}$.
The selective amplification of high-frequency transverse waves occurs in this manner.

\subsection{Origin of high-frequency waves}

In the previous subsection, we have shown that, via mode conversion, high-frequency longitudinal waves ($\nu_{HH}$) are converted to high-frequency transverse waves ($\zeta^{+}_{HH}$).
Therefore the origin of $\nu_{HH}$ is the key in our process.
To clarify it, using Fourier transformation, we decompose $\nu$ into low-frequency and high-frequency parts with cut-off frequency of $10 {\rm \ mHz}$.
In Figure \ref{fig:nu_space_time}, we show the space-time plots of the low-frequency part (a) and high-frequency part (b).
Panel (a) in Figure \ref{fig:nu_space_time} is similar with panel (b) in Figure \ref{fig:space_time} because the input frequency $f_0 = 5{\rm \ mHz}$ is lower than the cut-off frequency.
Panel (b) indicates that there are two origins of high-frequency longitudinal waves.
The first one lies in $z=0.5-1.0 {\rm \ Mm}$ and this corresponds to upward-wave origin, while the second one in $z=1.5-2.0 {\rm \ Mm}$ where downward waves are generated.
Considering the wave momentum conservation, only upward longitudinal waves are converted to upward transverse waves.
Figure \ref{fig:nu_space_time} indicates that high-frequency upward longitudinal waves are generated in $z=0.5-1.0 {\rm \ Mm}$, which is below the conversion region ($z=1.0-1.5 {\rm \ Mm}$).
This is consistent with Figure \ref{fig:zeta_freq}.
The possible physical mechanism of this high-frequency wave generation is wave steepening, because high-frequency components are always accompanied by low-frequency components.
However, we cannot rule out the possibility that the reflected waves play a role, because high-frequency upward waves are amplified after they collide with reflected high-frequency waves.

The physical processes in our simulation is summarized in Figure \ref{fig:sum}.
First, longitudinal waves from the photosphere steepen and high-frequency longitudinal waves are generated.
Such high-frequency longitudinal waves efficiently convert their mode to transverse waves by mode conversion, because the transmission rate is smaller for higher-frequency waves \citep{Schun06,Cally08}.
By the collision between longitudinal waves and transition region, (type-I) spicule is generated \citep{Hollw82b,Iijim15}, 
while high-frequency transverse waves probably appear as high-frequency spicule oscillations \citep{He00009,Okamo11}.

\subsection{Comparison with observation}

Transverse velocities at $z=2{\rm \ Mm}$ are shown in Figure \ref{fig:vt}.
The middle and right panels depict the low-frequency and high-frequency components, respectively, as decomposed by Fourier analysis with a cut-off period corresponding to $150 {\rm \ s}$,
which approximately equals the upper limit of the lifetime of spicules \citep{Perei12} observed by {\it Hinode}.
It should be noted that recent observations by {\it IRIS} \citep{DePon14} indicate a longer lifetime of spicules \citep{Skogs15}. Nevertheless, they are beyond the scope of the present study because the aim of this study includes a comparison with the {\it Hinode} observation.
The rms velocities of the high-frequency components correspond to $5.29 {\rm \ km \ s^{-1}}(v_x)$ and $4.97 {\rm \ km \ s^{-1}}(v_y)$, respectively.
In terms of amplitude, they correspond to $7-8 {\rm \ km \ s^{-1}}$, and this is consistent with the observed value \citep{Okamo11}.
Additionally, as shown in the lower left panel of Fig. \ref{fig:vt}, the typical period (duration time) of the high-frequency transverse waves is to $40 {\rm \ s}$.
It should be noted that the appearance of pulse-like fluctuation has a frequency of $f_0$.
The typical period of the pulse is also in accordance with  \citet{Okamo11}.
A natural interpretation is that this period represents the duration time of the mode conversion.
Specifically, the sound crossing time of the equipartition layer $\tau_{\rm MC}$ is given as follows:
\begin{align}
	\tau_{\rm MC} = \left. \frac{h}{c} \right|_{a=c} = \left. \frac{1}{c} \left[ \frac{d}{dz} \left( \frac{a^2}{c^2} \right) \right]^{-1} \right|_{a=c} \sim 40 {\rm \ s}.
\end{align}
This supports the interpretation.

We require a careful interpretation when we compare our results with observation.
Transverse waves in our simulation are a mixture of fast and Alfv\'en waves.
Alfv\'en waves are restricted such that they propagate along the field line while fast waves are refracted due to the high Alfv\'en speed gap between the chromosphere and the corona \citep{Rosen02,Bogda03}. 
The refraction is not considered in our simulation, and therefore our calculation potentially overestimates the amplitude of the transverse velocity.

\citet{Okamo11} argue that the energy flux estimated from observation is slightly lower than the amount required for coronal heating when the filling factor is considered.
The observed feature is consistent with high-frequency components of the simulation, and thus the fore-mentioned study could potentially omit the low-frequency wave contribution that is not observed by spicule oscillation.
In fact, in our calculation, the energy flux of the high-frequency components are a few times lower than the total energy flux.
Thus, the observed flux is sufficiently high when both the low-frequency waves \citep{DePon07a,McInt11,Thurg14} and high-frequency waves are considered.

\section{Summary} \label{sec:summary}

In this study, a numerical simulation was used to demonstrate that the longitudinal-to-transverse mode conversion is responsible for the observed spicule oscillation.
Figure \ref{fig:space_time} clearly shows direct evidence of mode conversion.
As a result of the mode conversion, high-frequency waves are selectively excited in the chromosphere (Fig. \ref{fig:zeta_freq}).
The behavior of high-frequency component is in agreement with the observed feature (Fig. \ref{fig:vt}).

Several wave generation and coupling processes exist in addition to mode conversion in the chromosphere and corona.
For example, \citet{Santa17} indicated that high-frequency waves are generated near the magnetic null point.
\citet{Marti17} showed that transverse waves are generated by the magnetic tension force induced by the ambipolar diffusion.
The multi-dimensional effect also induces the other wave coupling processes \citep{Hasan05,Hasan08,Antol15,Muraw15}.
Wave refraction is potentially important as discussed in the previous section, \citep{Rosen02,Bogda03}.
Additionally, an increasingly sophisticated treatment of radiation is potentially essential \citep{Hanst15,Iijim17,Marti17}.
In order to overcome these difficulties, it is necessary to perform a multi-dimensional radiation-magnetohydrodynamics (RMHD) simulation, and this will be explored in a future study.

The authors thank Yoshiaki Kato and Takeru K. Suzuki for insightful comments and fruitful discussions.
M. S. is grateful to Takayoshi Oba and Masashi Abe for advice on several topics related to observational studies.
M. S. is supported by Leading Graduate Course for Frontiers of Mathematical Sciences and Physics (FMSP) and  Grant-in-Aid for JSPS Fellows.
T. Y. is supported by JSPS KAKENHI Grant Number 15H03640.
Numerical calculations were partly performed on the PC cluster at the Center for Computational Astrophysics, National Astronomical Observatory of Japan.

\appendix
\section{Derivation of basic equations}

\begin{figure}[!t]
	\begin{center}
 	 \includegraphics[width=60mm]{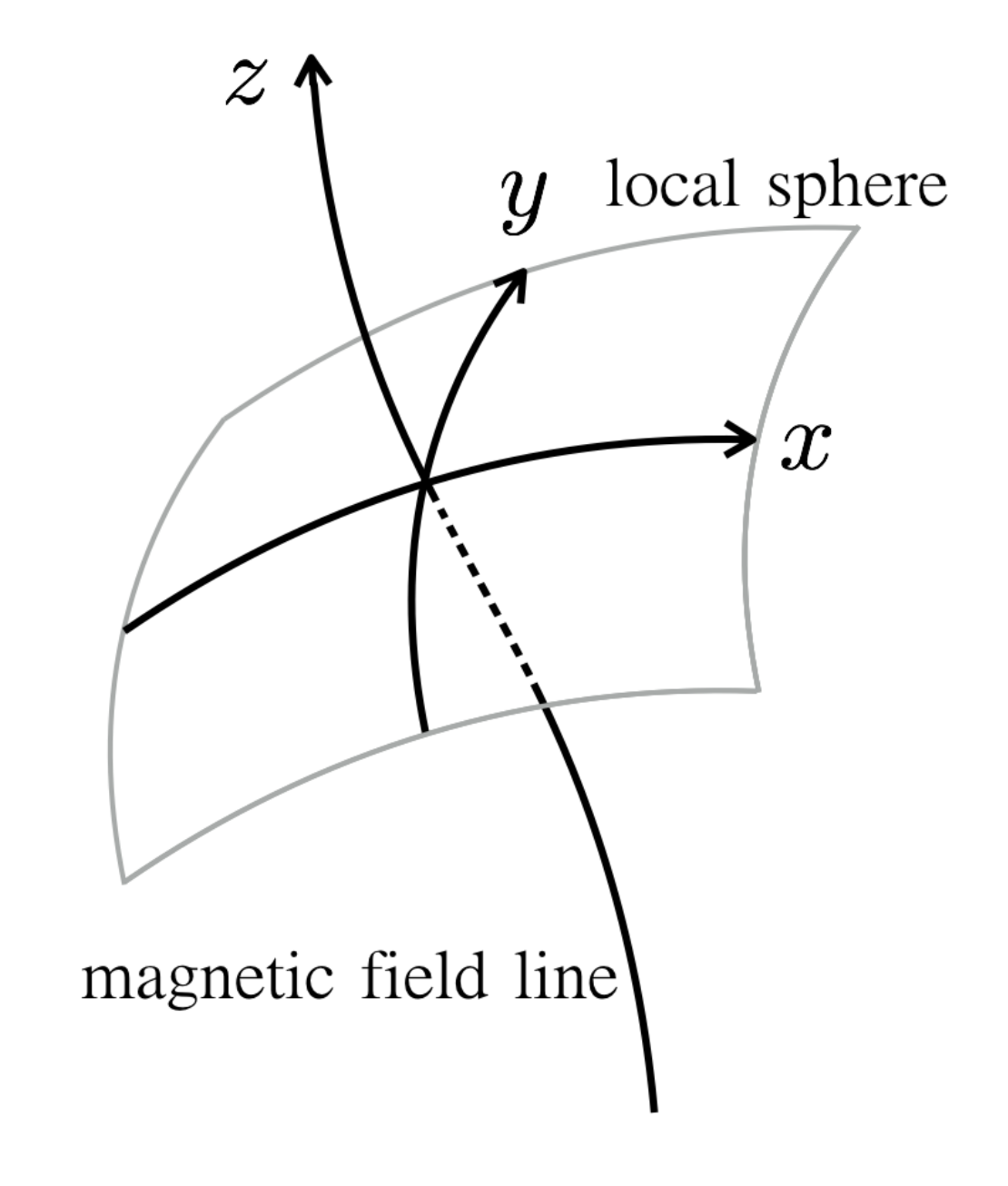}
  	\end{center}
  	\vspace{2em}
  	\caption{
	Schematic picture of the coordinate system in this study.
  	}
  	\label{fig:coordinate}
	\vspace{0em}
\end{figure}

The basic equations used in the study are derived in this appendix.
We set a coordinate, $z$, which is curved along the background magnetic field line and the other two coordinates, $x$ and $y$, which are orthogonal to $z$ axis such that the $xy$ plane is a local sphere.
The schematic picture is shown in Figure \ref{fig:coordinate}.
The coordinate curve that expands with the field line is capable of nonlinear evolution of Alfv\'en waves \citep{Hollw82a,Suzuk05,Antol08,Matsu10}.
We use several assumptions for the purpose of simplicity.
First, $x$ and $y$ are locally symmetric.
We consider a localized region near a thin flux tube in question, and the configuration of the flux tube is assumed as symmetric in $x$ and $y$ directions.
Second, the flux tube is sufficiently thin such that the curvature of $z$ axis is negligible when compared with $x$ and $y$ axes.
This is equivalent to the idea that the scale factors depend only on $z$.
Therefore, the $z$ derivatives of basis vectors are not considered.
Given these assumptions, the derivatives of basis vectors in this system are expressed as follows:
\begin{align}
	&\frac{\partial}{\partial x} \vec{e}_x = - \frac{1}{h} \vec{e}_z,  \ \ \frac{\partial}{\partial y} \vec{e}_x = 0, \ \ \ \ \ \ \ \ \frac{\partial}{\partial z} \vec{e}_x = 0, \nonumber \\
	&\frac{\partial}{\partial x} \vec{e}_y = 0,  \ \ \ \ \ \ \ \ \frac{\partial}{\partial y} \vec{e}_y = - \frac{1}{h}  \vec{e}_z, \ \ \frac{\partial}{\partial z} \vec{e}_y = 0, \label{eq:deri_base} \\
	&\frac{\partial}{\partial x} \vec{e}_z = \frac{1}{h}  \vec{e}_x, \ \ \ \ \frac{\partial}{\partial y} \vec{e}_z = \frac{1}{h}  \vec{e}_y,  \ \ \ \ \frac{\partial}{\partial z} \vec{e}_z = 0, \nonumber
\end{align}
where $h$ represents a scale factor determined by the structure of the flux tube.
One-dimensionality is assumed in which any physical variable depends only on $z$.
The solenoidal condition of the background magnetic field yields the following:
\begin{align}
	\left[  \vec{e}_x \frac{\partial}{\partial x} + \vec{e}_y \frac{\partial}{\partial y} + \vec{e}_z \frac{\partial}{\partial z} \right] \cdot \left( B_z \vec{e}_z \right) 
	= \frac{2}{h} B_z + \frac{d}{dz} B_z 
	= 0.
\end{align}
With respect to arbitrary $h$, it is possible to determine $A$ such that
\begin{align}
	\frac{1}{h} = \frac{1}{2A} \frac{dA}{dz}.
\end{align}
If we rewrite the solenoidal condition in terms of $A$, Eq. (\ref{eq:solenoidal}) is derived as follows:
\begin{align}
	\frac{B_z}{A} \frac{dA}{dz} + \frac{d}{dz} B_z  = \frac{1}{A} \frac{d}{dz} \left( A B_z \right) = 0.
\end{align}
$A B_z$ is conserved along the magnetic field line, and therefore $A$ denotes the cross section of the flux tube.
Given (\ref{eq:deri_base}), in terms of $A$, $\nabla \cdot \vec{f}$ and $\left( \vec{f} \cdot \nabla \right) \vec{g}$ are expressed as follows:
\begin{align}
	\nabla \cdot \vec{f} &= \frac{1}{A} \frac{\partial}{\partial z} \left( A f_z \right), \label{eq:divf} \\
	\left( \vec{f} \cdot \nabla \right) \vec{g} &= \left[ \frac{1}{2 A} \frac{dA}{dz} f_x g_z + f_z \frac{\partial}{\partial z} g_x \right] \vec{e}_x + \left[ \frac{1}{2 A} \frac{dA}{dz} f_y g_z + f_z \frac{\partial}{\partial z} g_y \right] \vec{e}_y +\left[ - \frac{1}{2 A} \frac{dA}{dz} \left(f_x g_x + f_y g_y \right) + f_z \frac{\partial}{\partial z} g_z \right] \vec{e}_z, \label{eq:fnablag}
\end{align}

The ideal MHD equations are as follows \citep{Pries14}:
\begin{align}
	&\frac{\partial }{\partial t} \rho + \nabla \cdot \left( \rho \vec{v} \right) = 0, \label{eq:ideal_mass} \\
	&\frac{\partial}{\partial t} \left( \rho \vec{v} \right) + \vec{v} \nabla \cdot \left( \rho \vec{v} \right) + \left( \rho \vec{v} \cdot \nabla \right) \vec{v} = - \nabla \left( p + \frac{\vec{B}^2}{8 \pi} \right) + \frac{1}{4 \pi} \left( \vec{B} \cdot \nabla \right) \vec{B}, \label{eq:ideal_eom} \\
	&\frac{\partial}{\partial t} \vec{B} + \left( \vec{v} \cdot \nabla \right) \vec{B} = - \left( \nabla \cdot \vec{v} \right) \vec{B} + \left( \vec{B} \cdot \nabla \right) \vec{v}, \\
	&\frac{\partial}{\partial t} e + \nabla \cdot \left( e \vec{v}  \right) + p \nabla \cdot \vec{v} = 0.
\end{align}
Given Eqs. (\ref{eq:divf}) and (\ref{eq:fnablag}), the MHD equations in our coordinate system are derived.
For example, the conservation of mass is obtained from Eqs. (\ref{eq:divf}) and (\ref{eq:ideal_mass}) as follows:
\begin{align}
	\frac{\partial}{\partial t} \rho + \frac{1}{A} \frac{\partial}{\partial z} \left( \rho v_z A \right) = 0,
\end{align}	
This is equivalent to Eq. (\ref{eq:mass}).
The equation of motion of $z$ component is given from Eqs.  (\ref{eq:divf}),  (\ref{eq:fnablag}) and  (\ref{eq:ideal_eom}) as follows:
\begin{align}
	\frac{\partial}{\partial t} \left( \rho v_z \right) + \frac{v_z}{A} \frac{\partial}{\partial z} \left( \rho v_z A \right) 
	- \frac{\rho}{2 A} \frac{d A}{d z} \left( v_x^2 + v_y^2 \right) + \rho v_z \frac{\partial}{\partial z} v_z = - \frac{\partial}{\partial z} \left( p + \frac{\vec{B}^2}{8 \pi} \right)
	- \frac{1}{2 A} \frac{d A}{d z} \left( \frac{B_x^2 + B_y^2}{4 \pi} \right) + \frac{B_z}{4 \pi} \frac{\partial}{\partial z} B_z. \label{eq:eomz_raw}
\end{align}
The following relations are used:
\begin{align}
	&\frac{v_z}{A} \frac{\partial}{\partial z} \left( \rho v_z A \right) + \rho v_z \frac{\partial}{\partial z} v_z = \frac{1}{A} \frac{\partial}{\partial z} \left( \rho v_z^2 A \right), \nonumber \\
	&\frac{\partial}{\partial z} p = \frac{1}{A} \frac{\partial}{\partial z} \left( p A \right) - p \frac{1}{A} \frac{dA}{dz}, \nonumber \\
	&\frac{\partial}{\partial z} \left( \frac{\vec{B}_\perp^2}{8 \pi} \right) + \frac{1}{2A} \frac{dA}{dz} \left( \frac{\vec{B}_\perp^2}{4 \pi} \right) = \frac{1}{A} \frac{\partial}{\partial z} \left( \frac{\vec{B}_\perp^2}{8 \pi} A \right), \nonumber
\end{align}
where $\vec{B}_\perp = B_x \vec{e}_x + B_y \vec{e}_y$, and thus Eq. (\ref{eq:eomz_raw}) yields the following:
\begin{align}
	\frac{\partial}{\partial t} \left( \rho v_z \right) + \frac{1}{A} \frac{\partial}{\partial z} \left[ \left( \rho v_z^2 + p + \frac{\vec{B}_\perp^2}{8 \pi} \right) A \right] 
	= \left( p + \frac{\rho \vec{v}_\perp^2}{2} \right) \frac{dA}{dz}.
\end{align}
The gravitational acceleration $-\rho g$ is added to the right-hand side, and the equation is multiplied by $A$ to obtain Eq. (\ref{eq:eomz}).
The other equations are derived in a similar manner.

\end{document}